\documentclass[12pt]{article}
\usepackage{graphics}
\usepackage{amsfonts}
\usepackage{amsmath}
\usepackage{amssymb}
\let\a=\alpha \let\b=\beta  \let\g=\gamma  \let\d=\delta \let\e=\varepsilon
      \let\k=\kappa \let\l=\lambda
             \let\p=\pi    \let\r=\rho
\let\s=\sigma \let\t=\tau    
   
 \let\D=\Delta   
         
\let\O=\Omega

\def\\{\hfill\break} \let\==\equiv

\let\0=\noindent

\def\nn{\nonumber}

\def\qed{\hfill\raise1pt\hbox{\vrule height5pt width5pt depth0pt}}

\def\be{\begin{equation}}
\def\ee{\end{equation}}
\def\bea{\begin{eqnarray}}\def\eea{\end{eqnarray}}

\begin{document}

\markright{Quasi-evaporating black holes...}

\title{Quasi-evaporating black holes and cold dark matter}

\author{Julien Larena$^*$ and Tony Rothman$^{*\dagger}$
\\[2mm]
{\small\it \thanks{julien.larena@gmail.com}}~ \it $^{\dag}$Dept. of Mathematics \& Applied Mathematics\\ University of Cape Town, \\
\it Rondebosch 7701, Cape, South Africa\\
{\small\it\thanks{trothman@princeton.edu}}~ \it Princeton University, \\
\it Princeton, NJ 08544}

\date{{\small   \LaTeX-ed \today}}

\maketitle

\begin{abstract}
Vilkovisky has claimed to have solved the black hole backreaction problem and finds that black holes
lose only ten percent of their mass to Hawking radiation before evaporation ceases.  We examine the
implications of this scenario for cold dark matter, assuming that primordial black holes are created during the reheating period after inflation.  The mass spectrum is expected to be dominated by 10-gram black holes.  Nucleosynthesis constraints and the requirement that the earth presently exist do not come close to ruling out such black holes as dark matter candidates.  They also evade the demand that the photon density produced by evaporating primordial black holes does not exceed the present cosmic radiation background by a factor of about one thousand.

 \vspace*{5mm} \noindent PACS: 4.70.-s, 4.70.Dy,95.35.+d,98.80.-k.,98.80.Ft

\end{abstract}

\section{Quasi-evaporating black holes}
\setcounter{equation}{0}\label{sec1}
\baselineskip 8 mm

In a recent series of  papers, Vilkovisky \cite{Vilk06a,Vilk06b,Vilk06c,Vilk08}
has claimed to have completely solved the black hole backreaction problem, which is the major unresolved issue regarding black hole physics since Hawking's discovery of black hole radiation thirty years ago \cite{Hawk75}.  Hawking's original calculation was semiclassical and ignored the effect of the black hole's evaporation on the spacetime metric itself; Hawking argued that any such backreaction should be negligible until the black hole approaches the Planck mass.  Vilkovisky maintains that if backreaction is properly taken into account, it exerts a negative feedback on the evaporation process, which ceases after the hole has lost only ten percent of its mass to thermal radiation. If the scenario is correct, then contrary to the accepted picture, primordial black holes of mass less than $10^{15}$ g will not have entirely disappeared by the present epoch of the universe and could conceivably provide the cold dark matter sought by cosmologists.  Vilkovisky's calculations are also semiclassical and do not assume any exotic physics.  Nevertheless, they have yet to receive full scrutiny by the physics community and it is premature to claim they are correct.  In this paper we merely propose to take his scenario seriously and examine the consequences.  We find that, indeed, black holes with mass less than $10^{15}$ grams can provide the $\O \approx$ .3 dark matter while safely evading basic observational constraints.  Such a conclusion may seem fairly obvious, since the black holes are evaporating only ten percent of their mass.  On the other hand, their evaporation profile does not at all resemble the usual one and it is worth verifying that this does not seriously alter the expected conclusions.  Furthermore, other phenomena, such as vortons\cite{CA00}, behave similarly at least insofar as that they are stabilized against contraction by the current they generate.   We do also feel that it is worthwhile to stimulate discussion of Vilkovisky's work, which if correct, may have provided the answer to two outstanding problems.\\

Carr et al. have provided several reviews of the standard picture of primordial black hole (pbh) formation processes and observational constraints on their mass density; see for example \cite{Carr05,Carr94,Carr75} and references therein.  Although the density of black holes with $M \gtrsim 10^{15}$ grams is constrained by a variety of techniques, including microlensing and $\gamma$-ray bursts, all constraints on holes with $M \lesssim 10^{15}$ grams are imposed by assuming that they interfere with established cosmological processes through their evaporation, which in the Hawking picture ends in an explosive phase.  In the Vilkovisky picture, by contrast, the temperature profile of the emitted radiation differs sharply from the usual one, being nearly flat until it drops to zero, and evaporation never reaches an explosive phase.

The Hawking temperature is the usual one,
\be
T = M_{pl}^2\frac{\k(u)}{2\p}
\ee
where $M_{pl}$ is the Planck mass, $\k$ is the surface gravity and $u \equiv t - r$ is the retarded time.  (We work in units with $c = G = k_{Boltzmann} = 1$.) However, in Vilkovisky's scenario the surface gravity is \emph{not} $\k = \frac1{4M}$.  His considerations of the radiation at infinity lead him to describe the metric in terms of two ``Bondi charges."  The first is ${\cal M}$, the Bondi mass, which is the mass remaining in a compact region after the escape of radiation (the ADM mass minus the mass of the radiation).  The second is a ``charge" $\cal Q$, which the black hole manufactures from the vacuum to protect itself from quantum evaporation.  We emphasize, however, that these two ``charges" are {\it not} merely assumed to exist but are a necessary consequence of the radiation metric.  They are in fact, the two leading coefficients of the metric expansion at infinity and are the only two that need to be specified.  Because in terms of these two coefficients, the asymptotic metric assumes the Reissner-Nordstr\"om form, the surface gravity becomes that for a Reissner-Nordstr\"om black hole:
 \be
\k = \frac{({\cal M}^2-{\cal Q}^2)^{1/2}}{[{\cal M}+({\cal M}^2-{\cal Q}^2)^{1/2}]^2}
\ee

The evolution of $\cal M$ and $\cal Q$ are governed by two coupled differential equations:
\bea
\frac{d{\cal M}}{du} &\approx & - \frac{M_{pl}^2}{48\p}\k^2\\
\frac{d{\cal Q}^2}{du} &=& \frac{M_{pl}^2}{24\p}\k .
\eea

We plot $\cal M,Q$ and $T$ in figure 1.  As one sees, $T$ is essentially flat but reaches a slight maximum of about
\be
T_{max} = \frac{M_{pl}^2}{8\pi M}, \label{TV}
\ee
before dropping abruptly to zero, at which point evaporation stops.  Here $M$ is the initial mass of the black hole.

\begin{figure}[htb] \vbox{\hfil\scalebox{.7}
{\includegraphics{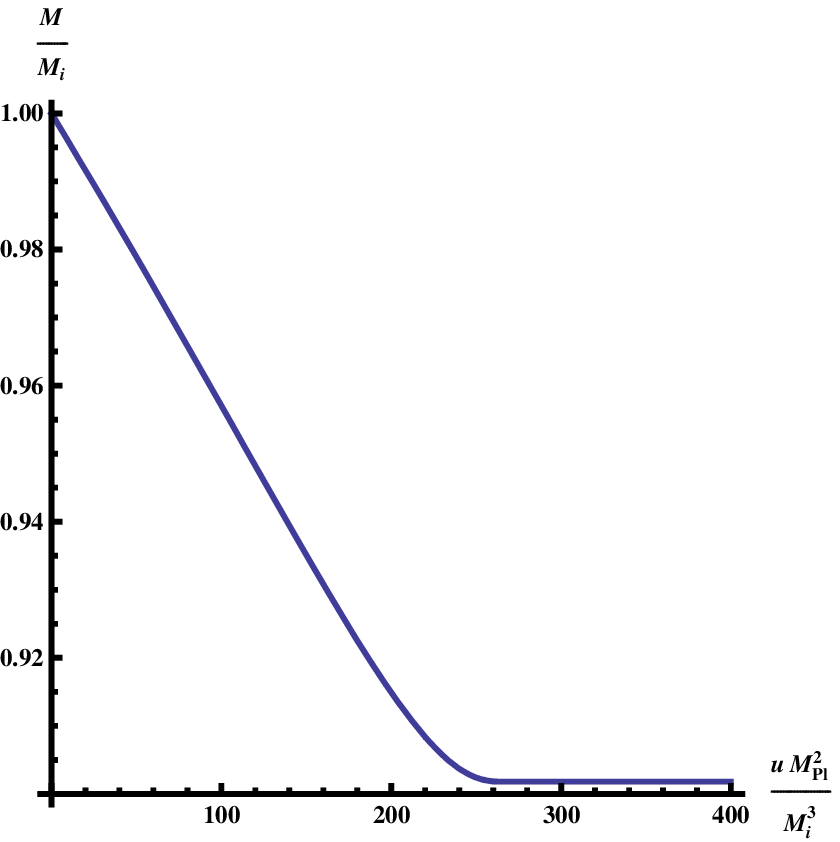}\includegraphics{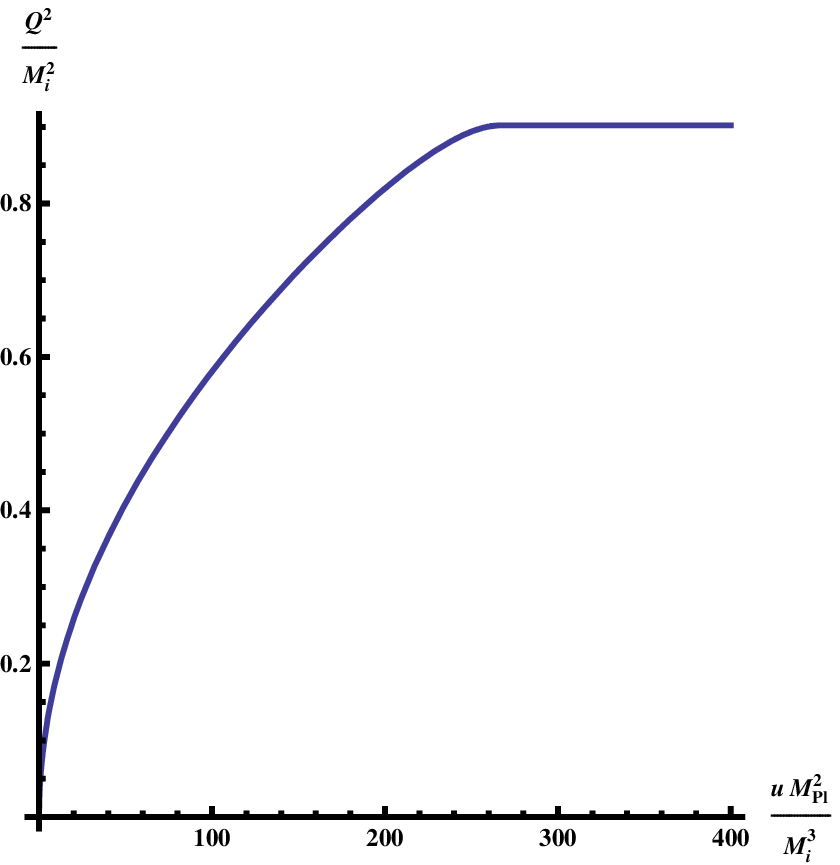}\includegraphics{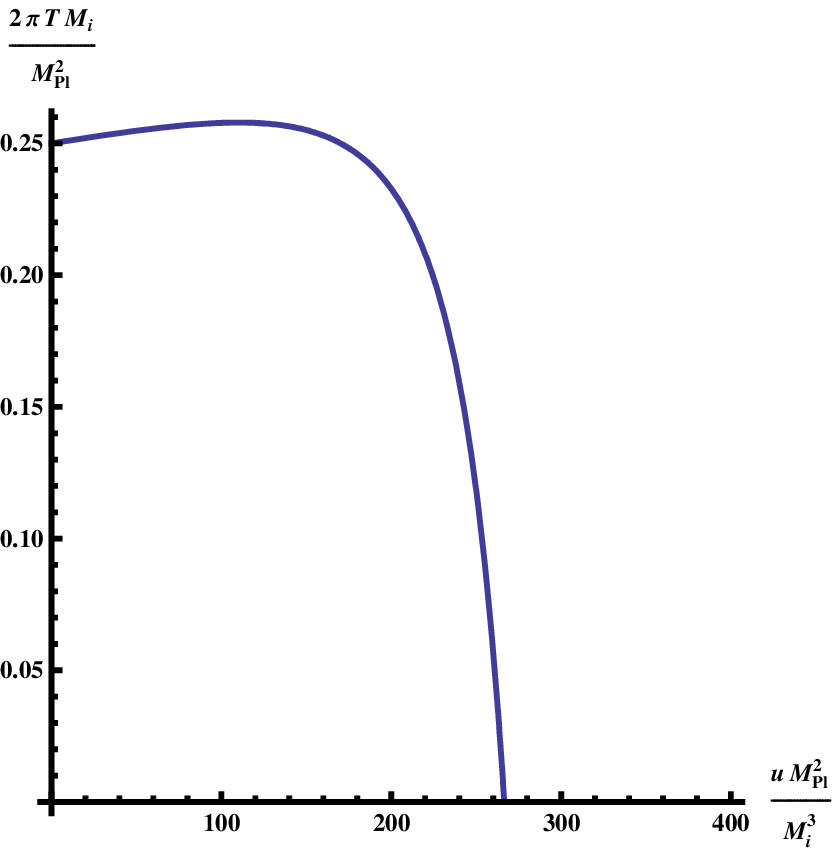}\hfil} {\caption{\footnotesize{
${\cal M},  {\cal Q}^2$ and $T$ plotted in terms of the initial mass vs. normalized retarded time $u$ for the Vilkovisky scenario.  Notice that the temperature is not proportional to $1/\cal M$ but is reasonably flat until it drops to zero and that $\cal M$ has dropped to about .9 of its initial value by the time evaporation ceases.}
 \label{figs}}}}
\end{figure}

In what follows we shall assume that
\be
T = T_{max} = constant = 4.8 \times 10^{20} \frac{M_{pl}}{M} \ \mathrm{MeV}\label{TV2}
\ee
or
\be
T \approx \frac{10^{15}}{M} \ \mathrm{MeV},  \label{TV3}
\ee
where $M$ is in grams.

The retarded time taken for evaporation is
\be
\D u \approx 96\p \frac{M^3}{M_{pl}^2},
\ee
where $u$ is evaluated at ${\cal I}^+$.  We assume that asymptotically $u$ becomes the cosmic time in a flat Friedmann-Lema\^ itre-Robertson-Walker universe, which in the matter-dominated case yields $\D u = \D t(1-3(t_0/\D t)^{2/3})$ for initial time $t_0$.
We see that for $\D t >> t_0$ we may take $\D u \approx \D t$.   Thus the cosmic time over which the hole is evaporating
\be
\t_{evap} \approx 96\p \frac{M^3}{M_{pl}^2} \approx 1.6 \times 10^{-27} M^3 \label{tevap}
\ee
for $M$ in grams.\footnote{As in Hawking's original work, Vilkovisky's calculation was carried out for a single scalar field.  If
$N$ species of particles are emitted with equal probability, the evaporation time will decrease by a factor $1/N$; however Page's results\cite{Page76}  show that emission is highly suppressed for particles with nonzero spin and thus only spin-zero need to be considered.  Unfortunately, we do not know the number of spin-0 species the energies of interest in this paper.  If $N \lesssim 10$, the results to be discussed go through unchanged.  If $N \gtrsim 10$, the limits to be discussed will necessarily be relaxed even further, because the integration time over pbh evaporation will be drastically reduced.\label{foot1}}

Equating $\t_{evap}$ with the age of the universe $t_u \approx 4.5 \times 10^{17}$ s, gives a  value,
\be
M_{u} \approx 6.6 \times 10^{14} \ \mathrm{grams}.
\ee
This is essentially the canonical $10^{15}$ gram black hole but in Vilkovisky's scenario it is not the mass of a hole that is evaporating during the current epoch; it is merely the mass of a hole whose evaporation is currently ceasing, by which time it will have lost only about ten percent of its mass, as can be seen in figure 1.  Thus, $M_{u}$ is the value above which black holes effectively do nothing whatsoever by the current age of the universe.  We expect that such a change should have drastic astrophysical consequences.

\section{PBH spectrum and total mass constraints}
\setcounter{equation}{0}\label{sec2}

We would like to check whether, given a reasonable pbh mass spectrum, there are any obvious reasons that pbhs of $M \lesssim 10^{17}$ grams (where femtolensing constraints kick in \cite{Carr05}) could not comprise the dark matter density $\O_{dm} \approx  .3$.  There are both observational and theoretical reasons to believe that the density fluctuation spectrum in the early universe should be scale-invariant, and in fact only in this case can pbhs have an extended mass spectrum.  Nevertheless, one expects any black holes formed before inflation to be diluted to negligible density by the exponential expansion and that therefore any relict pbhs would have been formed after inflation, presumably during the reheating phase.

In this case one faces a serious difficulty: Because reheating takes place as inflation ends, any wave mode perturbing the background at this time will have not entered the Hubble radius (``horizon") at the onset of inflation to be frozen in as it exited later on.  Consequently there is no clear reason for the perturbation spectrum to be scale invariant.  Carr\cite{Carr75,Carr05} gives the following expression for the probability that a spherical region of  mass $M$ forms a black hole:
\be
P(M) \sim \e \left(\frac{M}{M_*}\right)^{2/3 - n}\exp\left[-\frac{\g^2}{2\e^2}\left(\frac{M}{M_*}\right)^{2n-4/3}\right],\label{prob}
\ee
where $p = \g\r$ is the equation of state and $M_*$ is a fiducial mass.  The prefactor is actually the ratio $\d/\d_{max}$, where $\d \= dM/M $ is assumed to obey the power law $\d =  \e(M/M_*)^{-n}$ and $\d_{max} = (M/M_0)^{-2/3}$ results in maximal black hole formation.  In this formulation $n = 2/3$ represents the scale-invariant case and for $n > 2/3$ significant pbh production can take place only when $\e \sim 1$.  From Eq. (\ref{prob}) one can show that for $n = 2/3$ the number density of pbhs today with mass between $M$ and $M + dM$ should be
\be
dn(M) = \b\r_*\left(\frac{R_*}{R_0}\right)^3 M_*^{-2}\left(\frac{M}{M_*}\right)^{-5/2}dM, \label{npbhsi}
\ee
Here, $\r_*$ is the radiation density at the epoch when black holes of mass $M_*$ were formed, $R_0$ is the present-day scale factor and $R_*$ is the scale factor at the formation epoch, which we will assume to be the reheating period.  The factor $\b \= \e\exp\left(-\frac{\g^2}{2\e^2}\right)$ represents the fraction of regions of mass $M$ that collapses into black holes.

For $n \ne 2/3$ one can show with somewhat more difficulty that
\bea
dn(M) &\sim &  \r_*\left(\frac{R_*}{R_0}\right)^3M_*^{-2}\left(\frac{M}{M_*}\right)^{-11/6-n}
\exp\left[-\frac{\g^2}{2\e^2}\left(\frac{M}{M_*}\right)^{2n-4/3}\right]\nn\\
&\times&(\frac{2}{3}-n)\left[\frac{\g^2}{\e^2}\left(\frac{M}{M_*}\right)^{2n-4/3}-1\right]dM.  \label{npbh}
\eea

The $n = 2/3$ case is certainly the easiest spectrum to deal with and because it is a power-law, we expect it to be dominated by pbhs of $M = M_*$.  However, as mentioned, it is not obvious that pbhs forming as inflation ends should obey the scale-invariant spectrum.  From the usual Fourier decomposition of density perturbations\cite{KT90} one has $\d = \d\r/\r \propto k^{3/2}\d_k$ with $\d_k \propto k^{\a/2}$.  Evaluated at the wavelength corresponding to the mass, $\l \sim k^{-1} \sim M^{1/3}$ and $(\d\r/\r)_M\= dM/M \propto M^{-(1/2 + \a/6)}$.  Note that $\a = 1$ gives $\d \propto M^{-2/3}$, the Harrison-Zel'dovich spectrum.   From the Mukhanov equation\cite{Baum09}, one expects that any modes that do not cross the Hubble radius during inflation should have $\a=3$.  Equating exponents to the pbh-formation case, where  $\d \= (M/M_*)^{-n}$, gives $n = \a/6+1/2$.  For $\a = 3$, we have $n = 1$.

Notice from Eq.(\ref{npbh}), however, that for any $n \ne 2/3$, there is not only an exponential cutoff  but that
$dn(M)$ strictly vanishes when the last bracket equals zero.  Since one expects $\e \sim 1$ and $\g = 1/3$ for a hard equation of state, the $n=1$  case gives $M_{max} \sim 30 M_*$.  The exponential cutoff ensures that the number density will be negligible at an even smaller value of $M_{max}$.  From these considerations, it is reasonable to limit ourselves to the scale-invariant spectrum (for simplicity and ease of comparison) and to delta-function spectrum, in which all the pbh mass is concentrated in black holes of mass $M_*$.  As we will see, the results do not differ significantly.

For the scale-invariant case with a radiation equation of state, Eq. (\ref{npbhsi}) yields a total present mass density\cite{Carr75,Carr05}
\be
\O_{pbh}\r_{crit}=\r_*\left(\frac{R_*}{R_0}\right)^3\b M_*^{1/2}\int_{M_*}^{M_{max}}M^{-3/2}dM. \label{mspec1}
\ee
 Since $\r_{rad} \propto R^{-4}$ and $\r_{pbh} \propto R^{-3}$, Eq. (\ref{mspec1}) can be rewritten as
\bea
\O_{pbh}\r_{crit}&=&\O_{rad}\r_{crit}(1+z_*)\b M_*^{1/2}\int_{M_*}^{M_{max}}M^{-3/2}dM,\nonumber\\
&=&2\O_{rad}\r_{crit}(1+z_*)\b \left[1-\left(\frac{M_*}{M_{max}}\right)^{1/2}\right]\label{mspec2}
\eea
for formation redshift $z_*$. Note that placing constraints on the mass density of pbhs is equivalent to constraining $\b$.

In the standard scenario, one typically cuts the spectrum off at the lower limit $M_* \approx 10^{15}$ grams because smaller holes would have evaporated by today. Taking the upper limit to be $\infty$ gives the total mass density above the cutoff.  In Vilkovisky's scenario, as already explained, smaller pbhs are still present.  If we take $M_*$ to be the minimum mass of pbhs formed after inflation, then we expect it to   be roughly the horizon mass at the time of formation\cite{Carr05}, or in a radiation-dominated universe
\be
M_* = M_{pl}\left(\frac{T_{rh}}{T_{pl}}\right)^{-2}
\ee
where $T_{rh}$ is the reheating temperature. Since $T_{rh} \approx 10^{16}$ GeV and $T_{pl} \approx 10^{19}$ Gev, we have $M_* \approx 10^6 M_{pl} \approx 10$ g.  Then, because $M_* <<< M_{max}$, Eq. (\ref{mspec2}) becomes simply
\be
\O_{pbh}\approx \O_{rad}(1+z_*)\b ,\label{mspec3}
\ee
independent of the upper limit  $M_{max}$. For $\O_{pbh} \le .3$, $\O_{rad} \sim 10^{-4}$ and $z_* \sim 10^{29}$, the above yields $\b \sim 10^{-26}$.\\

To obtain a delta-function spectrum, we  replace the $M_*^{-2}dM$ in Eq. (\ref{npbh}) by $M_*^{-1}\d(M-M_*)dM$, where now $\d$ is a Dirac delta function, and integrate over $M$, giving
\be
n = \b\r_*\left(\frac{R_*}{R_0}\right)^3M_*^{-1}.  \label{npbhd}
\ee
In this case $\b$ is a different functional form of $\e$ than the scale-invariant spectrum but the physical interpretation remains the same, the fraction of the volume of space that has collapsed into black holes.  With the approximations we have made, Eq. (\ref{mspec3}) remains unchanged for the delta-function spectrum, and the limit on $\b$ the same.

\section{Entropy and annihilation  constraints}
\setcounter{equation}{0}\label{sec3}

 The above limit was set merely by requiring the total mass density in primordial black holes not exceed the known dark matter density and it is not immediately obvious that this value for $\beta$ is free of observational contradictions.  Specifically, it does not take into account pbh evaporation, which can be expected to affect the cosmic background photon density.  The strongest constraints\cite{Carr94} in the standard model assume that $10^{15}$ g pbhs are currently evaporating in the $\g$-region. Requiring that density of $\g$'s from pbhs be less than that of the background, pushes $\b$ down to about $10^{-28}$.  We now perform a similar analysis for the Vilkovisky scenario, although here there is a significant difference: Since black holes $\lesssim 10^{15}$ g continue to evaporate to the present at an approximately constant rate, for the scale-invariant spectrum we must integrate the photon emission over the entire history of the universe.  (From Eq. (\ref{TV2}) it is clear that the temperature of small black holes can be high enough to evaporate massive particles.  In the present analysis, we assume that only massless particles, which we refer to as ``photons," are emitted; this will only tighten the constraints. See also footnote \ref{foot1}.)

For the scale-invariant case, if $.1M$ worth of photons is evaporated per hole, then  Eq. (\ref{npbhsi}) allows us to write for the photon energy density due to pbhs at emission
\be
d\r_{\g e}  = .1\b\r_*\left(\frac{T_e}{T_*}\right)^3 M_*^{1/2}M^{-3/2}dM. \label{drg}
\ee
Here the subscript $e$ denotes ``at emission" and the temperature $T \propto R^{-1}$ refers to the background temperature.   This equation assumes that all pbhs are continuously evaporating, but after a time $\t_{evap}$ given by Eq. (\ref{tevap}), a black hole shuts off and no longer contributes to  $\r_\g$.  Inverting Eq. (\ref{tevap}) therefore gives in a lower-limit cutoff for $M$ as a function of $t$.  If $M_*$ is the smallest allowable cutoff and we take $T \propto t^{-2/3}$ over most of the history of the universe, then $M_{cut} = M_*(T_{\t *}/T_e)^{1/2}$, where  $T_{\t*}$ is the temperature at the time when a hole of mass $M_*$ ceases to evaporate.  Given a formation time $t_*$ for pbhs of mass $M_*$, we have $t_*/\t_* = M_{pl}^2/(96\p M_*^2)$. Assuming that formation takes place in a radiation-dominated phase, then  $T_{\t *}/T_* = (t_*/\t_*)^{1/2}$ and thus $T_{\t *} = (M_{pl}/\sqrt{ 96\p} M_*)T_*$.  Consequently,
\[
M_{cut} = \left(\frac{M_{pl}}{\sqrt {96\p}}\right)^{1/2}M_*^{1/2}\left(\frac{T_*}{T_e}\right)^{1/2} \approx 10^{-3} M_*^{1/2}\left(\frac{T_*}{T_e}\right)^{1/2},
\]
where the last expression is for $M_{pl}$ in grams.

 Now integrating Eq. (\ref{drg}) between $M_{cut}$ and $M$ and retaining only the lower limit  gives
\be
\r_{\g e} \approx .2 \times 10^{3/2}\b\r_*M_*^{1/4}\left(\frac{T_e}{T_*}\right)^{13/4}.
\ee
Redshifted to the present background temperature $T_0$, $\r_{\g e}$ becomes $\r_{\g 0} = \r_{\g e} (T_0/T_e)^4$ or
\be
\r_{\g 0} \approx .2\times 10^{3/2}\b\r_*M_*^{1/4}\left(\frac{T_0^4}{T_*^{13/4}}\right)T_e^{-3/4}.
\ee
This can be regarded as the integral of $d\r_{\g 0}$  evaluated at $T_e$ and $T_0$, or
\be
\r_{\g 0} \approx .2\times 10^{3/2}\b\r_*M_*^{1/4}\left(\frac{T_0}{T_*}\right)^{13/4} = .2\times 10^{3/2}\b\r_0M_*^{1/4}\left(\frac{T_*}{T_0}\right)^{3/4},
\ee
where we have once more discarded the upper limit and have set $\r_* = \r_{0} (T_*/T_0)^4$, where $\r_0$ is the current radiation density.

 Since $T_*/T_o \sim 10^{29}$ and $M_* \approx 10$ g, then $\r_{\g 0}  \sim .3 \times 10^{23.5}\b\r_0$. If we require that $\r_{\g 0} \lesssim \r_0$, then $\b \lesssim 3\times  10^{-23.5}$.  Because this is significantly larger than the $\b \sim 10^{-26}$ found by requiring  $\O_{pbh} \le .3$, 10-g pbhs are not ruled out as dark matter candidates at the level of this calculation for the scale-invariant spectrum.\\

 For the delta-function spectrum, we take $\r_{\g e} = \r_{pbh} \times (\mathrm{number \ of\ photon's\ per\ hole}) = .1 M\r_{pbh}/T_{\g h}$, where $T_{\g h}$ is the energy of a $\g$ emitted by the black hole.  Then  from Eq. (\ref{npbhd})
 \be
 \r_{\g e} = .1\b\r_*\left(\frac{T_e}{T_*}\right)^3\frac{M_*}{T_{\g h}},
 \ee
or in terms of $\r_0$ and redshifted to the present as above,
\be
\r_{\g 0} = .1\b\r_0\left(\frac{T_*}{T_e}\right)\left(\frac{M_*}{T_{\g h}}\right). \label{rd}
\ee
 We approximate $T_e$ by $T_{\t *}$, the background temperature at evaporation; as calculated above $T_{\t *} \sim 10^{-10} T_{pl}$.  From Eq. (\ref{TV3}), $T_{\g h} \sim 10^{-8} T_{pl} > > T_{\t *}$ but photons from 10-gram black holes will thermalize on a timescale far shorter than the expansion timescale and merely become an indistinguishable part of the background radiation density.  Thus instead of $T_{\g h}$ we use $T_{\t *}$ in Eq. ({\ref{rd}) and find that $\b \sim 10^{-22}$, which is reasonably consistent with the scale-invariant calculation.

We note that the constraint $\r_{\g 0} \lesssim \r_0$ is the weakest that one can apply, but seems to be the appropriate one for these models.  As just mentioned, any photon emitted before decoupling will thermalize.  From Eq. (\ref{tevap}) only photons from pbhs of $M \gtrsim 10^{13}$ g will be emitted after decoupling and not thermalize.  Eq. (\ref{TV2}) shows that the energy of such photons is $\sim 100$ MeV, or about 100 KeV when redshifted to the present.  The x-ray background in this part of the spectrum is indeed only $\sim 10^{-5}$ of the CMB background, but with the scale-invariant mass spectrum, holes in the $10^{13}$-g range form a fraction only $\sim 10^{-30}$ of the total and therefore can make only a negligible contribution to the x-ray background.  For the delta-function spectrum such holes do not even exist.  Nevertheless, any constraint that is a factor of $\sim 10^{3}$ tighter than the one we have imposed will rule out these models.\\

In searching for further constraints we should consider the following:   Because pbhs of $M\sim 10$ g dominate either mass spectrum, for $\O_{pbh} \sim 1$ we must have $n \sim 10^{-30}$ cm$^{-3}$.  Assuming the same density enhancement within the galaxy above the average density of the universe implies $n \sim 10^{-25}$.  In that case collisions with the earth would be expected on a timescale $\t \sim (\s n v)^{-1} \sim 10^{-3}\mathrm{(c/v)\ s}$. For $v \sim 200 $ km/s, consistent with galactic rotation curves, we have $\t \sim 1$ s.  Thus, over its lifetime, the earth has experienced $\sim 10^{17}$ black hole collisions. This is a somewhat alarming prospect; one is  therefore prompted to investigate the effects of accretion resulting from such interactions.

In the simplest model\cite{GM08,Bloch09} (nonrelativistic, good in any number of dimensions), a sub-atomic-size black hole swallows anything that comes within a collision cross section:
\be
\frac{dM}{dt} \approx \p r_n^2\r v,  \label{accrete}
\ee
where $r_n \sim 10^{-13}$ cm is the radius of the nucleon, $\r$ is the average density of the medium and $v$ is the black hole velocity.  This accretion law is valid for $r_{bh} < r_n$ and with $v=$ constant yields trivially
\be
M_f \sim \p r_n^2 \r v t. \label{accrete2}
\ee
We may require that the earth presently exists.  Since $M_e \sim 10^{28}$ g, if each of the $10^{17}$ pbhs grew to $10^{11}$ g, this would presumably be an indication that the earth was annihilated by accretion.  The shortest accretion timescale $\t_{acc}$ will be given by the maximum $v$.  Taking $v$ to be the impact velocity $\sim 10^{7}$cm/s, and $\r_e \sim 5$ g/cm$^3$, Eq. (\ref{accrete2}) however yields $\t_{acc} \sim 10^{29}$ s.\footnote {The above accretion law can be written $dM/dx \approx \s\r$ or $dM \sim m_n dx/\l$, which states that the black hole picks up a nucleon's worth of mass every mean free path $\l$, independent of velocity.  Thus for a black hole to accrete $M_f \sim 10^{11}$ g requires $10^{11}$ nucleons, or about $10^{35}$ cm.}

More sophisticated analysis will not lower this number into the realm of measurability, regardless of spectrum.  Hence, the continued existence of the earth puts no limits on the density of 10 g pbhs.  Because $\t_{acc}$ scales as $M/\s \propto M^{1/3}$ for an astronomical body of mass $M$, neither will observing the potential destruction of asteroids significantly improve the annihilation constraint.

\section{Nucleosynthesis Constraints}
\setcounter{equation}{0}\label{sec4}

Big Bang nucleosynthesis (BBN) has also been used to limit pbh density (see \cite{Nov79,Lind80,RM81,KY00} and references therein). All BBN constraints on pbhs involve black holes of mass $\sim 10^{9-10}$ g, which are evaporating during helium production at 1 s$ \lesssim t \lesssim 100$ s.  Various processes can be considered, which either spallate the helium as it is being produced and overproduce deuterium, or destroy the deuterium itself (by photodisassociation). For such scenarios the delta-function spectrum of 10-gram black holes is clearly inapplicable, since all the black holes will have ceased  evaporation long before BBN, and because we have assumed that the emitted photons are thermalized, by BBN they can no longer be energetic enough to have any particular effect.

For the scale-invariant mass spectrum, one should put in a lower mass cutoff, as we did in \S \ref{sec3}, but one appropriate for a radiation-dominated universe, where $T \propto t^{-1/2}$:
\[
M_{cut} \approx 10^{-3} M_*^{1/2}\left(\frac{T_*}{T_e}\right)^{2/3}.
\]
However this cutoff implies that the only black holes continuing to evaporate at BBN are those with mass $\gtrsim 10^{10}$ g, as expected.  Thus we are effectively constraining black holes of same mass range as the earlier studies and  may merely cite their results, all of which give $\b \lesssim 10^{-21-22}$.  These figures should be relaxed by an order of magnitude to reflect the fact that in the Vilkovisky scenario only $.1 M$ is released as radiation. If we regard $10^9$-g black holes as the tail-end of the power-law spectrum, then a single value of $\b \lesssim 10^{-20}$ applies across the entire mass range.  This is a very weak limit compared to that obtained by constraining the present-day density of pbh photons and does nothing to rule out 10-g pbhs as dark matter candidates.\\

Thus, to conclude, the standard methods employed to limit pbh density place no limits on the black holes in the Vilkovisky picture.  To some extent this was expected, since quasi-evaporating black holes lose only ten percent of their masses to radiation.  In any case, although one might have theoretical prejudices against primordial black holes, such prejudices are not sufficient to rule them out as cold dark matter candidates.\\

{\bf Acknowledgements}
We would like to thank Alex Hamilton and Jean-Phillipe Uzan for helpful discussions and Steve Boughn for having a look at the manuscript.  JL acknowledges support from the Claude Leon Foundation, South Africa.

{\small
\end{document}